\documentclass[12pt]{article}
\usepackage[margin=1in]{geometry}
\usepackage{setspace}
\usepackage[T1]{fontenc}
\usepackage{fancyhdr}
\usepackage[USenglish]{babel}
\usepackage[utf8]{inputenc}
\usepackage[T1]{fontenc}
\usepackage{epsfig}
\usepackage{wrapfig}
\usepackage{color}

\usepackage[vflt]{floatflt}
\usepackage{longtable}
\usepackage{caption}
\usepackage{amssymb,amsmath}
\usepackage{longtable}
\usepackage{amssymb}
\usepackage{amsmath, xspace}
\usepackage{graphics,graphicx} 
\usepackage{rotating}
\usepackage{multirow}
\usepackage{subfigure}

\usepackage[dvipsnames,svgnames,x11names]{xcolor}
\usepackage[colorlinks=true, linktocpage, linkcolor={blue!40!black}, citecolor={blue!40!black}, urlcolor={blue!50!black}]{hyperref}
\usepackage[all]{hypcap} 

\usepackage[hang,flushmargin]{footmisc}

\usepackage{natbib}
\newcommand{\ignore}[1]{}

\definecolor{DarkGreen}{rgb}{0.0, 0.3, 0.0}
\definecolor{purple}{rgb}{0.5, 0.0, 0.5}
\definecolor{red}{rgb}{1, 0.0, 0.0}
\definecolor{green}{rgb}{0, 1.0, 0.0}

\newcommand{\expf}[1]{{{\rm e}^{#1}}}
 
\newcommand{\Moon}{$M_{Moon}$}                          

\newcommand{\apjl}{\rm ApJL}
\newcommand{\apjs}{\rm ApJS}

\newcommand{\aap}{\rm A$\&$A}

\def\ao{{\it Appl.~Opt.}}           

\def\3he{$^3{\rm He}$}
\def\arcdeg{\hbox{$^\circ$}}

\def\lsim{\mathrel{\lower2.5pt\vbox{\lineskip=0pt\baselineskip=0pt
           \hbox{$<$}\hbox{$\sim$}}}}

\def\gsim{\mathrel{\lower2.5pt\vbox{\lineskip=0pt\baselineskip=0pt
           \hbox{$>$}\hbox{$\sim$}}}}

\begin{document}

\thispagestyle{empty} 
\singlespacing

\begin{center} 

{\bf \large {\Large Wide Bandwidth Considerations for ALMA Band 2}} \\

\vspace{0.3cm}
{\noindent 
Tony Mroczkowski\footnote{European Southern Observatory, Karl-Schwarzschild-Strasse 2, Garching 85748, Germany}, 
 Carlos De Breuck$^1$,
 Ciska Kemper$^1$,
 Neil Phillips$^1$,
 Gary Fuller\footnote{Jodrell Bank Centre for Astrophysics \& UK ALMA Regional Centre Node, School of Physics and Astronomy, The
University of Manchester, Oxford Road, Manchester, M13 9PL, UK}, \\
 Maite Beltr\'an\footnote{INAF-Osservatorio Astrofisico di Arcetri, Largo E. Fermi 5, 50125, Firenze, Italy},
 Robert Laing\footnote{Square Kilometre Array Organisation, Jodrell Bank Observatory, Lower Withington Macclesfield Cheshire, SK11 9DL, UK\\\rule{\textwidth}{0.4pt}},
 Gianni Marconi$^1$,
 Leonardo Testi$^1$,
 Pavel Yagoubov$^1$,
 Danielle George$^2$, 
 William McGenn$^2$
 }
\end{center}

\singlespacing
\pagenumbering{arabic}

\setcounter{tocdepth}{2}
\setcounter{secnumdepth}{4}
\setcounter{footnote}{0}

\section*{Executive Summary}

One of the main considerations in the ALMA Development Roadmap\footnote{See \url{http://library.nrao.edu/public/memos/alma/main/memo612.pdf}, or see \url{http://www.eso.org/sci/facilities/alma/announcements/20180712-alma-development-roadmap.pdf} for a more graphical version.} for the future of operations beyond 2030 is to at least double its on-sky instantaneous bandwidth capabilities. Thanks to the technological innovations of the past two decades, we can now produce wider bandwidth high-performance radio frequency (RF) and intermediate frequency (IF) bandwidth receivers than were then foreseen at the time of the original ALMA specifications. In several cases, the band edges set by technology at that time are also no longer relevant. In this memo, we look into the scientific advantages of beginning with Band 2 when implementing such wideband technologies.  

The Band 2 receiver system will be the last of the original ALMA bands, completing ALMA's coverage of the atmospheric windows from 35--950 GHz, and is not yet covered by any other ALMA receiver. New receiver designs covering and significantly extending the original ALMA Band 2 frequency range (67-90~GHz) can now implement these technologies.  
We explore the scientific and operational advantages of a wide RF band, sideband-separating (2SB) receiver covering the full 67--116~GHz atmospheric window. The proposed receiver will cover an IF bandwidth of at least 4-16~GHz, with a goal of 4-18~GHz.  This wide IF design anticipates potential upgrades to the backend electronics that go beyond the minimal doubling of ALMA's bandwidth.
Here we consider the case that it achieves this 4-18~GHz IF band.
Deploying a wide RF band system on ALMA will not only open the currently unavailable 67-84~GHz range, but it could serve as a smooth deployment of a wideband system covering the full 67-116~GHz window, with reduced impact on science operations.

In addition to technological goals, the ALMA Development Roadmap provides 3 new key science drivers for ALMA, to probe:
1) the Origins of Galaxies, 2) the Origins of Chemical Complexity, and 3) the Origins of Planets.  
In this memo, we describe how the wide RF Band 2 system can help achieve these goals, enabling several high-profile science programmes to be executed uniquely or more effectively than with separate systems, requiring an overall much lower array time and achieving more consistent calibration accuracy: contiguous broad-band spectral surveys, measurements of deuterated line ratios, and more generally fractionation studies, improved continuum measurements (also necessary for reliable line flux measurements), simultaneous broad-band observations of transient phenomena, and improved bandwidth for 3~mm very long baseline interferometry (VLBI). We provide a comparative assessment of the effectiveness of ALMA to carry out a variety of different science goals when equipped with a wide RF Band 2 system.

Finally, we note that state of the art, high performance receivers covering a frequency range similar to the wide RF system are already in use at other (sub)mm observatories. This new Band 2 receiver could serve as one of the key upgrades that is contributing to close the wide bandwidth capability gap between ALMA and other facilities, and demonstrates the competitiveness, feasibility and effectiveness of this wideband concept.
The memo specifically uses a metric for comparison of the wide and narrow Band 2 cases, as defined by the ALMA Integrated Science team. The case presented therefore is entirely predicated on the condition that the wideband receiver cartridge meet this benchmark.


\tableofcontents

\section{Introduction}

\begin{figure}[tbh!]
\centering
\includegraphics[width=0.8\textwidth]{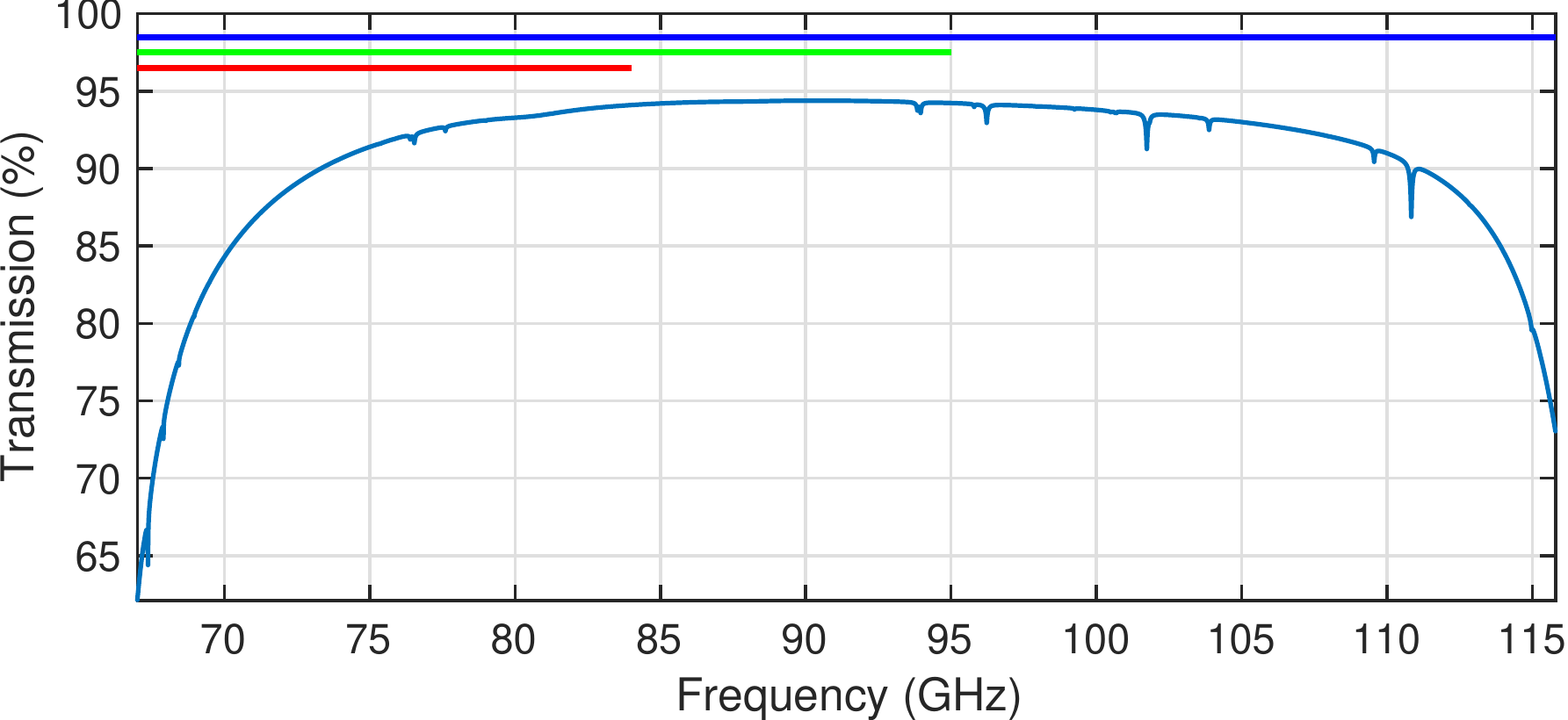}
\includegraphics[width=0.8\textwidth]{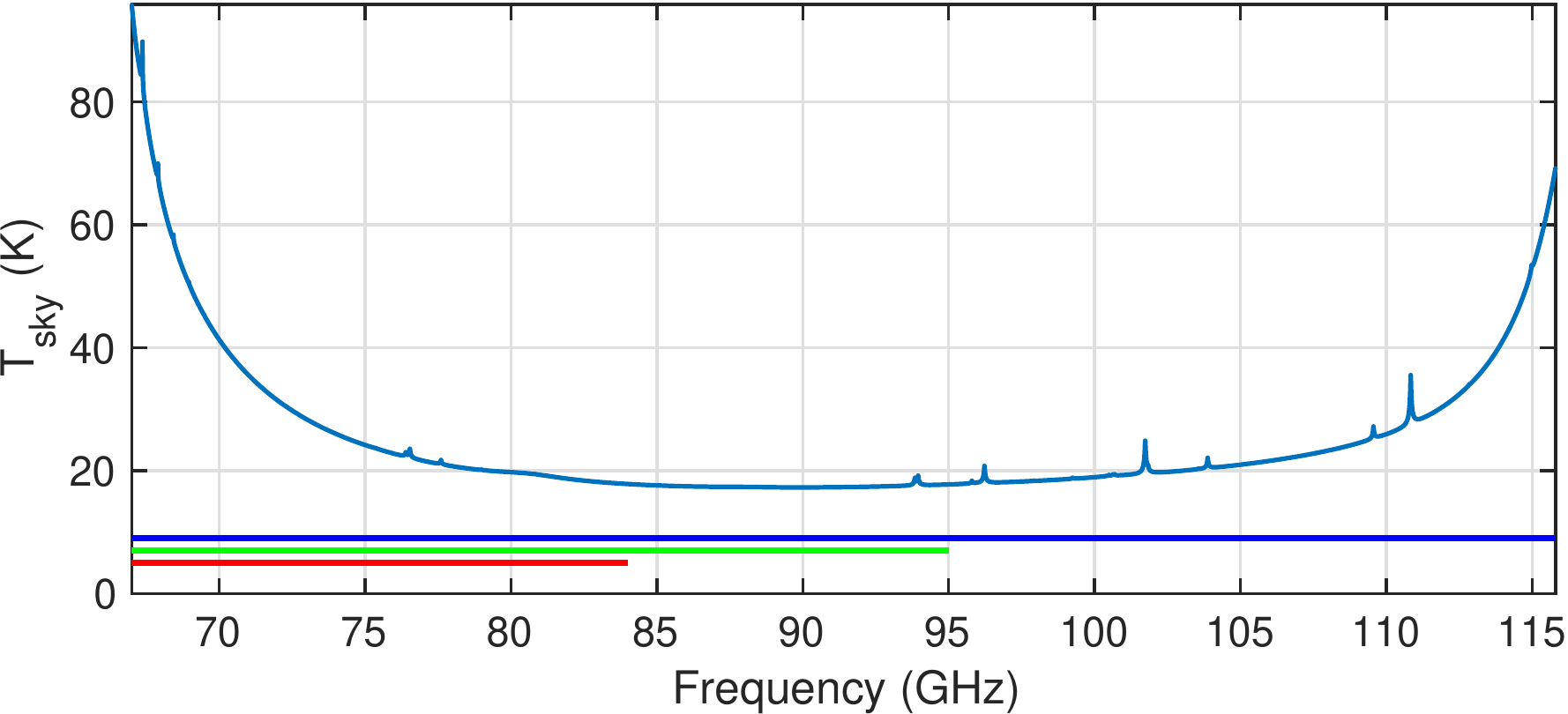}
\caption{\footnotesize Percent transmission (upper panel) and sky noise temperature ($T_{\rm sky}$ in Kelvin, lower panel) for observations taken at an elevation angle 60$^\circ$ in 7$^{\rm th}$ octile (PWV=5.178~mm) condition, using the {\tt am} model of \citet{paine_scott_2018_1193646}.
The red horizontal bar denotes the portion of the atmospheric window not yet covered by the current ALMA Band 3 receivers (i.e. 67-84~GHz).  
The green and blue horizontal bars show the coverage of the proposed ``narrow RF Band 2'' and ``wide RF Band 2'' receivers, respectively.  
The original ALMA Band 2 definition (67-90~GHz) covers less than half of the wide RF Band 2 range.
}
\label{fig:sky}
\end{figure}

With the publication of the ALMA Development Roadmap, the goal of broadening the receiver intermediate frequency (IF) bandwidth by \emph{at least} a factor two has been put forward as one of the two top priorities for ALMA.
The frequency ranges for the ALMA system were defined two decades ago, based on the then current and expected evolution of receiver technology performance and the science drivers (MMA Memo \#213, May 1998).\footnote{\url{http://legacy.nrao.edu/alma/memos/html-memos/alma213/memo213.html}} The 67--90~GHz frequency range, Band 2, is the final band of the original ALMA band specification yet to be approved for construction.  
 Two prototype sideband separating (2SB) designs for this band have been developed, going beyond the original ALMA Band 2 system definition that it be single sideband (SSB).  Both designs take advantage of bandwidth improvements made possible by technological innovations since the original ALMA band definitions. One design, described by \citet{Yagoubov2018}, aims to cover the complete 67--116~GHz atmospheric window, bracketed by the 60 and 118~GHz atmospheric oxygen lines, while the other design would cover 67--95~GHz, only partially overlapping with the 84--116~GHz range covered by the current Band 3 receivers (see Figure~\ref{fig:sky}).  Hereafter, we refer to the receiver spanning the on-sky radio frequency (RF) band of 67--116~GHz as the ``wide RF Band 2'', and we refer to the receiver spanning an RF band 67--95~GHz as the ``narrow RF Band 2.''

A wide RF and IF Band 2 system is extremely relevant to the three main scientific priorities defined in the ALMA Development Roadmap: the origins of galaxies; the origins of chemical complexity; and the origins of planets. 
While \cite{Beltran2015} and \cite{Fuller2016} address in depth the science that a receiver covering 67--116~GHz can deliver, 
we here consider the potential observational advantages of a wide-band system with RF bandwidth 67-116~GHz and IF bandwidth 4-18~GHz.  We note that the EA-ALMA partners are now considering 4-20~GHz IF, a technology upgrade that will benefit essentially all ALMA science, as noted in the EA-ALMA Development Workshop summary (A.\ Gonzalez, December 2018).\footnote{The workshop website is available here: \url{https://alma-intweb.mtk.nao.ac.jp/~diono/meetings/EA_ALMA_WS_2018/}.}

Several key advantages are uncovered by realising a wider bandwidth: observing time effectiveness, redshift searches and broad spectral scan capabilities, ability to probe spectral features spread broadly in frequency space in one spectral setup, ability to leverage instantaneous spectral indices, better sensitivity and imaging dynamic range for continuum observations, and potentially a significant improvement in the sensitivity and bandwidth of ALMA Band 3. 
Band 3 currently plays an essential operation role for ALMA, particularly for calibration and observations in suboptimal weather (e.g.\ $\approx 30\%$ of ALMA’s science integration time was spent in Band 3 in the last 4 cycles). The wide RF Band 2 presents an opportunity to extend this capacity.
Several of the capabilities discussed – particularly where only 1 spectral setup is required – will especially benefit transient and time-critical observations, in addition to improving ALMA’s overall observational efficiency.

As optimisation of the wideband receiver design and of the individual wideband components can lead to a trade-off in sensitivity versus the narrow band case, we also consider the impacts of such a trade-off on narrow band (mainly single line) observations.  A metric for comparison, which is discussed below, was recently put forward after careful consideration of the overall impact on science operations.  We estimate the quantitative impact of this trade-off on the down selection between receiver cartridges and the noise temperatures of the low noise amplifiers (LNAs) in Section~\ref{sec:metric} below.

\section{Comparison metric for wide versus narrow RF band}
\label{sec:metric}

\begin{figure}[tbh!]
\centering
\includegraphics[width=0.8\textwidth]{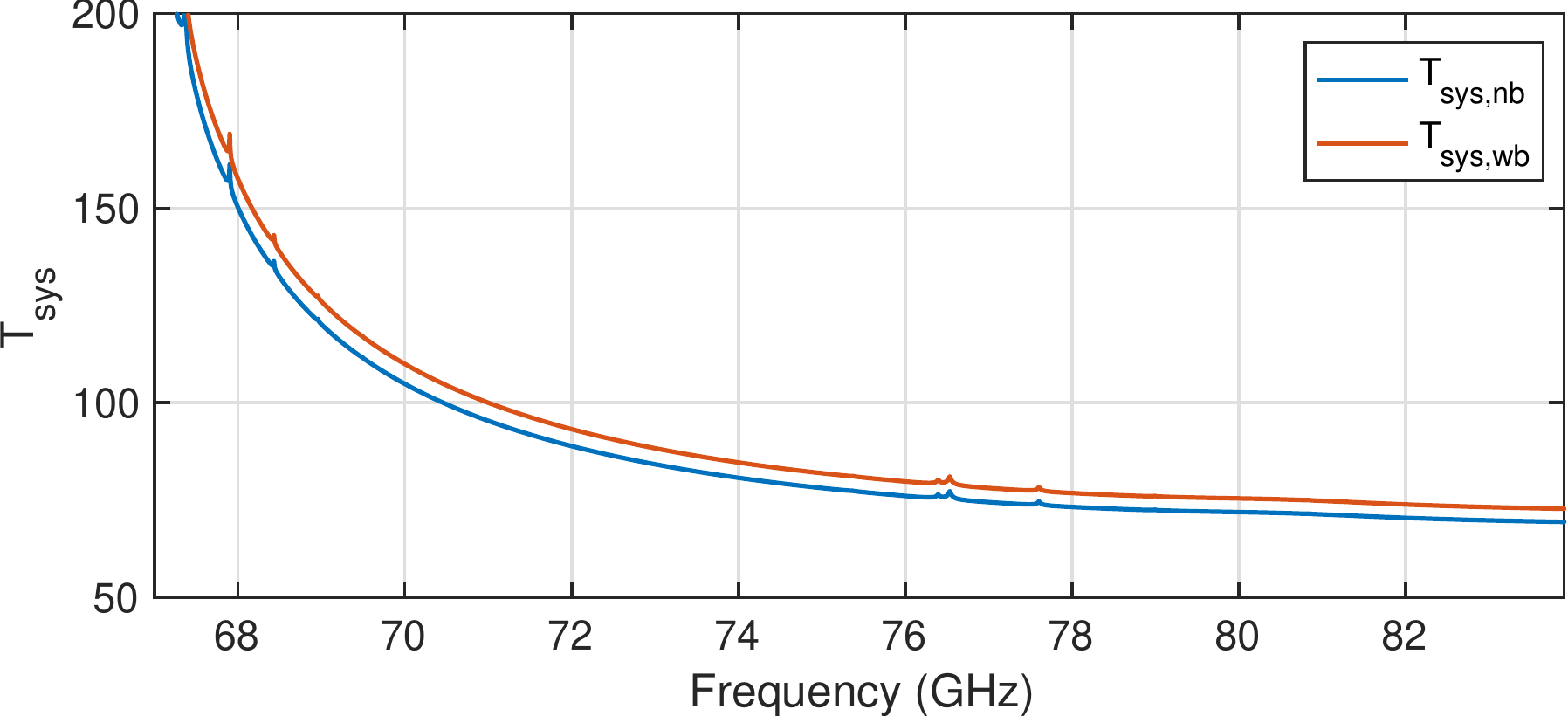}
\caption{\footnotesize System temperature assuming the same conditions as we do throughout this memo (see Figure~\ref{fig:sky}.
The lines show the narrow band $T_{\rm sys,nb}$ value resulting from assuming $T_{\rm rx}=30~\rm K$ in the 67-84~GHz range, and the new $T_{\rm sys,wb} = \sqrt{1.1} \, T_{\rm sys,nb}$.
}
\label{fig:tsys}
\end{figure}

One can expect a narrow RF band receiver to be better optimised for sensitivity than a wide RF band receiver using similar technology, so the question becomes {\it What trade off in instantaneous sensitivity are we willing to accept in order to achieve a wide bandwidth?}.
Our figure of merit for these comparisons is relative observational speed (i.e.\ the speed to complete the science goal and reach the same sensitivity for imaging, line detection, or a spectral scan for each system considered). 
This sensitivity metric was proposed in an internal memo (private communication; Carpenter, De Breuck, Iono, \& Wooten), and states that a 10\% increase in average integration time in the 67-84~GHz window -- the part of Band 2 not covered by the current Band 3 cartridges --  will be tolerated in exchange for improved bandwidth. 
The memo recommends adopting an elevation angle of 60$^\circ$ above the horizon and 7th octile atmospheric conditions for the comparison.
For a given frequency and bandwidth, this implies that a frequency-averaged system temperature of the wideband receiver in the 67-84~GHz can be the square root of 10\% higher (i.e.\ $T_{\rm sys,wb} \leq \sqrt{1.1} \, T_{\rm sys,nb}$; see Figure \ref{fig:tsys}).

Using Equation 6 of ALMA Memo \#602, ``ALMA Sensitivity Metric for Science Sustainability Projects,'' for the ALMA-wide standard definition of the system temperature $T_{\rm sys}$, which accounts for optical efficiency, atmospheric opacity, and a number of fundamental noise contributions (e.g.\ the quantum noise limit and image sideband contributions),   
we have\footnote{\url{http://library.nrao.edu/public/memos/alma/memo602.pdf}} 
\begin{equation}
T_{\rm sys} = \frac{1}{\eta_{\rm eff} \, \expf{-\tau_0 \sec{\theta_z}}}\left[
(1+g)(T_{\rm rx}+\frac{h\nu}{2k}) 
+ \eta_{\rm eff}(T_{\rm sky,s}+g T_{\rm sky,i})  
+ (1-\eta_{\rm eff})(T_{\rm amb,s}+g T_{\rm amb,i})
\right]
\label{eq:tsys}
\end{equation}
where 
\begin{itemize}

\item $\eta_{\rm eff}$ is the forward efficiency of the telescope+receiver as a function of frequency. We assume $\eta_{\rm eff} = 0.95$ as in ALMA Memo \#602.

\item $\expf{-\tau_0 \sec{\theta_z}}$ is the opacity of the atmosphere at zenith angle $z$ (i.e.\ $\theta_z = 90^\circ - \phi$ for the elevation angle $\phi$).

\item $g$ is the sideband gain ratio. Since Band 2 will be sideband separating, we assume $g=0$ as recommended in ALMA Memo \#602.

\item $(T_{\rm rx} + \frac{h\nu}{2k})$ is the receiver noise, where the factor $\frac{h\nu}{2k}$ arises from half-photon fluctuations, relevant for coherent/heterodyne receivers, and the energy of a photon with frequency $\nu$ is $E_\gamma = h\nu$.

\item $T_{\rm sky,s}$ and $T_{\rm sky,i}$ are the atmospheric (sky) noise contributions in the signal and image sidebands.  Again, the image sideband noise contribution $T_{\rm sky,i}=0$ for sideband separating receivers like Band 2.

\item $T_{\rm amb,s}$ and $T_{\rm amb,i}$ is the ambient (or ground) temperature, assuming optical spillover of the beam is onto the warm surroundings (compared to the sky). We assume an ambient temperature of $T_{\rm amb}=260~K$, which is generally appropriate for the ALMA site.  

\end{itemize}

For $g=0$, this reduces to
\begin{equation}
T_{\rm sys} = \frac{1}{\eta_{\rm eff} \, \expf{-\tau_0 \sec{\theta_z}}}\left[T_{\rm rx}+\frac{h\nu}{2k} 
+ \eta_{\rm eff} T_{\rm sky}  
+ (1-\eta_{\rm eff}) T_{\rm amb}
\right]
\label{eq:tsys_simp}
\end{equation}

The observational speed scales with bandwidth and inversely as the square of the system temperature.  
In this memo, we use the full $T_{\rm sys}$ calculation (Equation~\ref{eq:tsys}) following ALMA memo \#602. We compute $T_{\rm sky}$ and opacity using the {\tt am} atmospheric model \citep{paine_scott_2018_1193646} in Bands 2\&3, which is in relatively good agreement with Juan Pardo’s {\tt atm} code used by ALMA (e.g.\ in CASA and TelCal).  The transmission and $T_{\rm sky}$ for Bands 2\&3, assuming 7th octile conditions (PWV=5.178~mm, elevation angle 60$^\circ$), are shown in Figure \ref{fig:sky}.

Solving for the factor $a$ by which $T_{\rm rx}$ can be relaxed, assuming $T_{\rm sys,wb} \leq \sqrt{1.1} \, T_{\rm sys,nb}$, we have
\begin{equation}
a = \left[(\eta_{\rm eff} \, \expf{-\tau_0 \sec{\theta_z}}) T_{\rm sys}  - \frac{h\nu}{2k} 
- \eta_{\rm eff} T_{\rm sky}  
- (1-\eta_{\rm eff})T_{\rm amb}\right]/T_{\rm rx}
\label{eq:trx_relaxed}
\end{equation}

\begin{figure}[tbh!]
\centering
\includegraphics[width=0.8\textwidth]{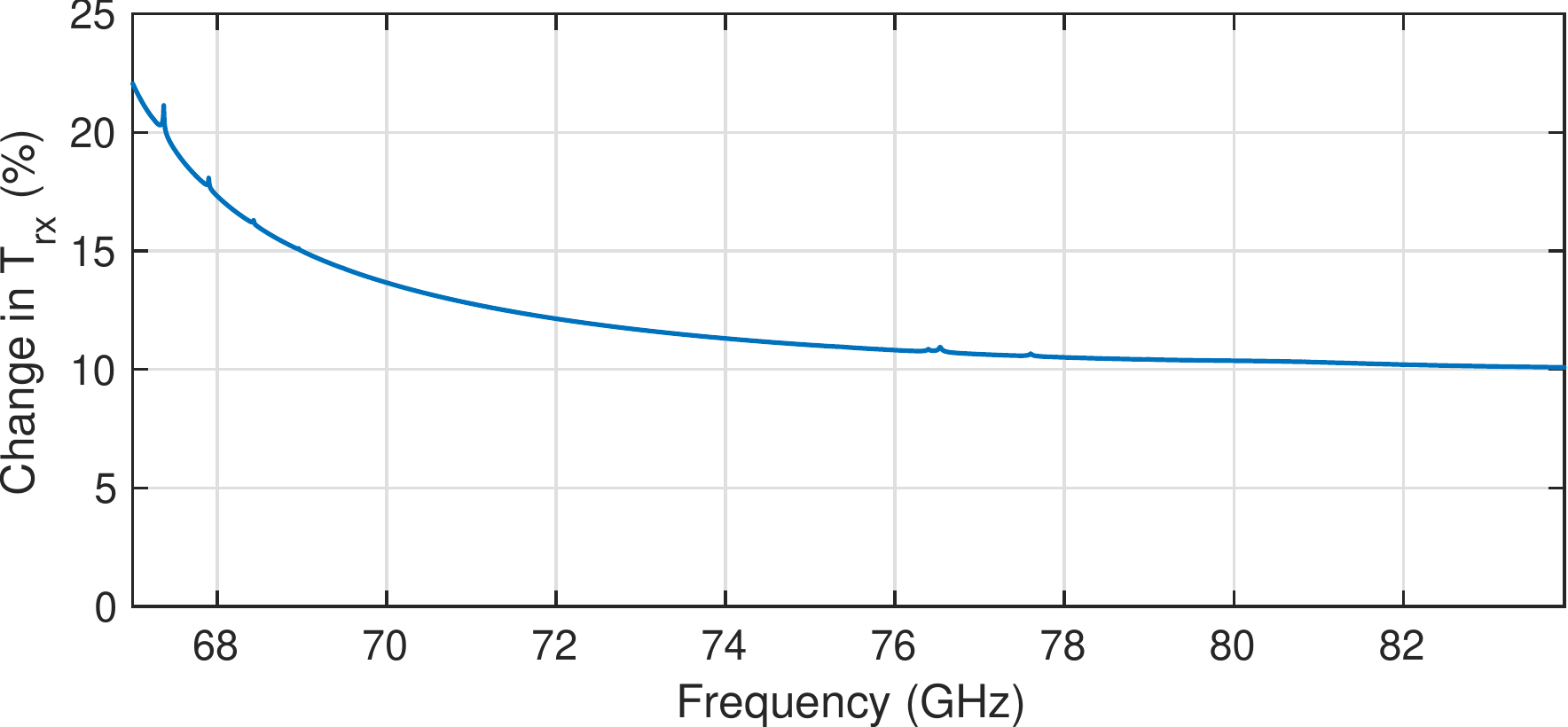}
\caption{\footnotesize Percentage change in the value of $T_{\rm rx}$ in the wide versus narrow RF band cases for our metric ($T_{\rm sys,wb} = \sqrt{1.1} \, T_{\rm sys,nb}$), defined at the beginning of \S \ref{sec:metric}. Here we assume the same atmospheric conditions as we do throughout this memo (see Figure~\ref{fig:sky}).
}
\label{fig:trx}
\end{figure}

The percentage by which $T_{\rm rx,wb}$ can be relaxed is plotted in Figure \ref{fig:trx}.
As mentioned in the introduction, the transmission and noise temperature at the edges of the 67-116~GHz window is dominated by oxygen lines which significantly increase $T_{\rm sys}$ at the band edges regardless of $T_{\rm rx}$ or PWV (see Figure \ref{fig:sky}).  Accounting for both opacity and $T_{\rm sky}$, it is clear from Figure \ref{fig:trx} that $T_{\rm rx}$ can be relaxed more than 5\% ($\sqrt{1.1}$), particularly at the edges of the band.  Instead, the wideband receiver noise specification allows 10-20\% higher $T_{\rm rx}$ than that for the narrow band case.\footnote{We clarify here that, when discussing $T_{\rm rx}$, we refer to the overall receiver cartridge noise temperature, which should be optimised as a whole to avoid losses due to mismatched components.  When comparing component performances, it is useful to consider the cascaded noise formula, $T_{\rm rx} \approx T_1 + T_2/(G_1) + T_3/(G_1 G_2) + \ldots$, where losses in the passive optical components imply $G_1<1$.
We note this model is a simplification, and there can be uncertainties due to impedance mismatches, so while giving useful first-order guidance on how the noise scales, this estimate does not replace the need for empirical characterisation of the components in the full system.}

The metric defined above assumes that the narrow RF receiver would be better optimised for deep integrations on single lines, which was one of the fundamental science drivers at the time of construction. For such observations, the IF bandwidth or the other sideband does not matter. Of course, a lower $T_{\rm rx}$ for such a narrow RF compared to a wide RF receiver still needs to be demonstrated. It is difficult to predict the evolution of the demand for such deep single-line observations compared to wideband observations, but a 10\% loss in observing time was considered an acceptable compromise. 
In Appendix \ref{sec:appendixb}, we provide evidence that the level of performance required to meet the wideband metric may be achievable with state-of-the-art receiver technology today.

The advantage of IF bandwidths up to 14~GHz can only be fully exploited in Band 2 if it is combined in a wider RF bandwidth to allow an optimal tuning flexibility throughout the entire atmospheric window. As we will see below, in the optimal case, the 67-116~GHz range can be covered in just 2 spectral setups. Potential upgrades of the current Band 3 receivers are also subject of an ongoing study, but these are not expected to cover the 67-84 GHz RF range. One alternative Band 2 complement to such an upgraded Band 3 receiver would be a fixed tuning receiver covering 67-84 GHz RF with an IF bandwidth of 17 GHz. Such a very wide IF bandwidth is stretching the current technology, and to be efficient for spectral surveys, would also require a wide IF Band 3 receiver. As the latter is beyond the scope of this memo, we here concentrate only on the wide RF Band 2 option.

\section{Impact on Spectral Line Studies}
\label{sec:narrowband}

A significant fraction of ALMA observations\footnote{Roughly one-third in Cycle 6 according to estimates from J.\ Carpenter.} seek deep observations of a single line at a known frequency and over a narrow bandwidth.  The impact of the relaxed wideband receiver noise temperature is immediately obvious from the metric as defined above: 
a 10\% increase in the average integration time for observations in the 67-84~GHz window will be tolerated in exchange for a receiver that yields improved wideband performance over the full Band 2+3 atmospheric window.

\section{Wideband Observational Cases}
\label{sec:wideband}

\subsection{Spectral scans across Bands 2 and 3}

\begin{figure}[htb]
\centering
\includegraphics[width=0.99\textwidth]{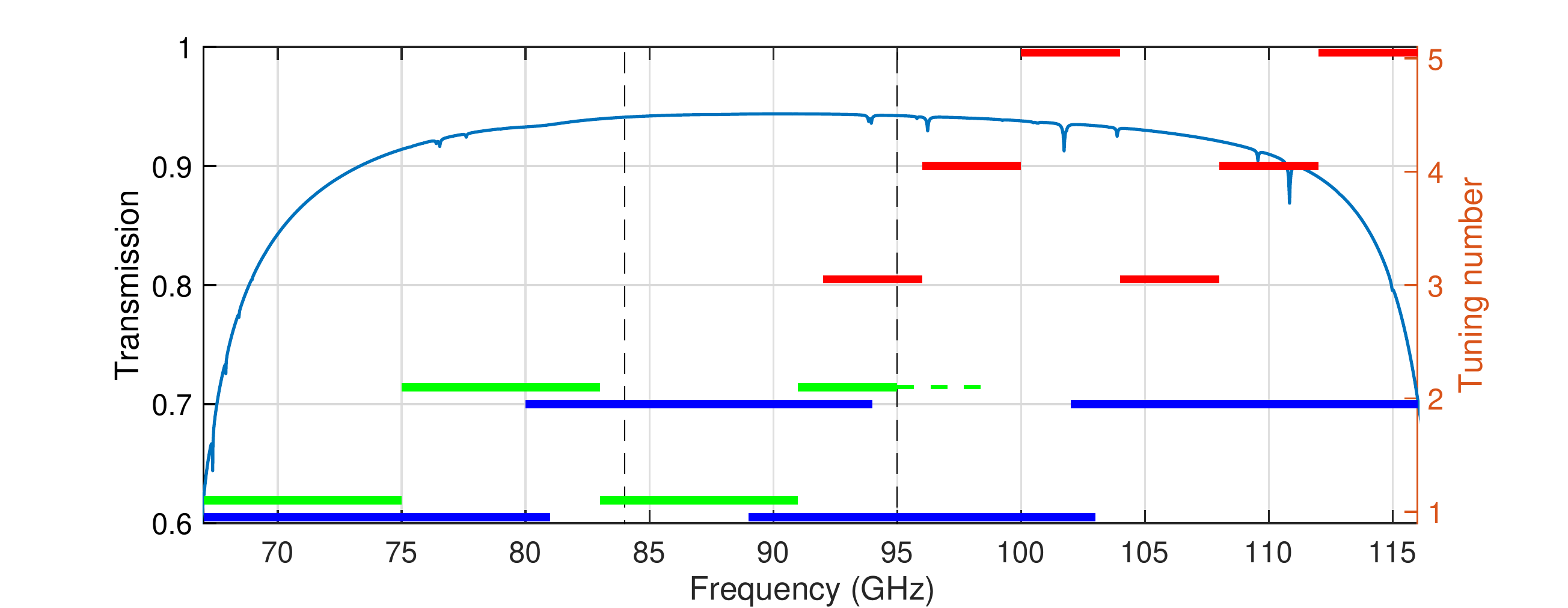}
\caption{
\footnotesize 
Minimum number of spectral tunings to cover the entire Band 2 \& 3 atmospheric window.  
The left axis shows the atmospheric transmission at 60 degree elevation assuming PWV$=$5.186~mm (7th octile conditions on the Chajnantor Plateau).
 The right axis shows the tuning number.  The vertical dashed lines indicate the lower edge of the current Band 3 receiver (84~GHz) and the upper edge of the narrow RF Band 2 design (95~GHz). 
 A wide RF Band 2 receiver with a 4-18 GHz IF band requires only 2 tunings (blue) to cover the entire window,  while the combination of the existing Band 3 (red) and narrow RF Band 2 (green) receivers would require 5 tunings.  
}
\label{fig:tunings}
\end{figure}

Three out of ten currently approved ALMA large programmes are using spectral scans: ASPECS (2016.1.00324), ALCHEMI (2017.1.00161) and the ALMA lensing cluster survey (2018.1.00035). 
In addition, a range of high-z, extragalactic and galactic full or partial Band 3 spectral scan programmes have been approved in all ALMA Cycles. We consider here how a wide RF band upgrade could improve the efficiencies of future high-redshift and other spectral surveys in achieving approximately uniform depth and complete spectral coverage of the 67-116~GHz atmospheric window.  After upgrades to the digitisers and correlator, a combined wide RF band receiver with 2$\times$14~GHz IF outputs will be able to cover this window in only 2 tunings with minimal overlap between the tunings (Figure~\ref{fig:tunings}).

To cover the full atmospheric window using a separate narrow RF band system along with the current Band 3 receivers requires a minimum of 5 tunings, which is also demonstrated in Figure \ref{fig:tunings} (green and red tunings).  In the example, the second Band 2 tuning has an unavoidable redundancy with the first tuning in Band 3; even worse, some of the tuning range is outside of the narrow RF bandwidth (dashed green lines), producing no useful data.  Three tunings are required for the original Band 3 to cover the 95-116~GHz portion of the spectrum not covered by the narrow RF Band 2 design. While the details of the narrow RF Band 2 plus Band 3 spectral setup could be adjusted, there is no clear way to reduce the number of tunings required.

For science goals requiring more than one spectral setup, the relative integration time also scales with number of observations required.  We note that reduction in the number of observations required will also simplify the calibration and data reduction, expedite quality assurance level 2 (QA2), and lower the risk that a project requiring multiple executions will not be completed.  The use of one receiver to cover the full wavelength range may also lead to more uniform sensitivity, and in the case when the full 4-18~GHz IF is ingested, we benefit from an overlap of the spectral window edges (better agreement of the flux scales between different tunings).

We can conclude that for observations requiring contiguous spectral coverage, the wide, widely separated sidebands of ALMA are best matched by building receivers with increased RF and IF bandwidth. 

\subsection{Widely Spaced Transitions}

The second key science driver in the 2030 ALMA Development Roadmap is the origin of chemical complexity.  This science goal can be met by probing the widely spaced molecular transitions of what are thought to be the building blocks of life.

After future upgrades to the digitisers and correlator, the broad IF bandwidth proposed for Band 2 will allow ALMA to probe simultaneously any two line transitions separated by any arbitrary amount from 0-36~GHz in a single observation (i.e.\ either in one 14~GHz wide sideband, or split between the 2 sidebands if more than $>8$~GHz apart).  This is particularly important in the lower ALMA bands, where many of the fundamental molecular transitions lie, and thus the spectrum of cold molecular gas is less crowded.  The current Band 3 receivers can only probe separations less than 4~GHz (i.e.\ in one sideband) and/or spanning 8-16~GHz (by placing SPWs in the upper and lower sidebands).  The relatively narrow bandwidth compared to the gap of 8~GHz between the upper and lower sidebands (USB and LSB) consequently leaves gaps for lines separated by 4-8 or $>$16~GHz that simply cannot be probed in one spectral setup with the current Band 3 (even after upgrades to the correlator and digitisers).  By requiring twice the spectral setups, the observation time required doubles (taking into account also the need of duplicating several calibrations); additionally, the uncertainty in the absolute gain calibration will be increased and will affect the line ratio measurements.  Logistically, requiring multiple executions also increases the risk of not completing an observation, or completing them in different antenna configurations (and hence spatial resolutions) if they are separated significantly in time.  The wide RF band, on the other hand, can simultaneously probe separations up to 14~GHz in a single sideband, or from 8-36~GHz by choosing SPWs across the 2 sidebands.  This will increase the observational flexibility for a wide range of science goals.

\begin{figure}[htb]
\centering
\includegraphics[width=0.999\textwidth]{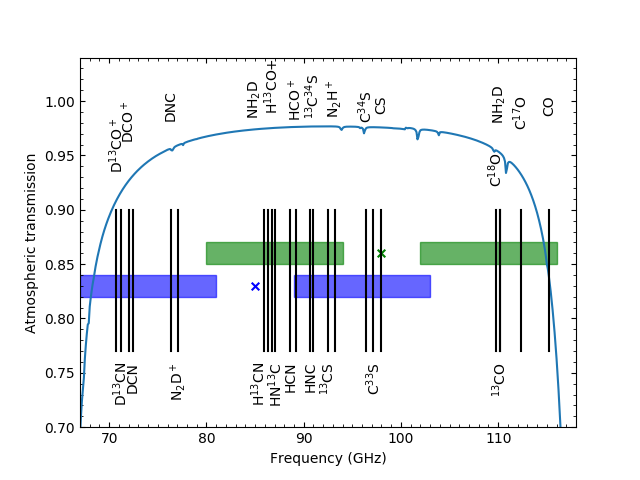}
\caption{\footnotesize Two spectral setups with the wide RF+IF Band 2 would cover the entire 67-116 GHz atmospheric window. See the main text for a discussion of interesting combinations of potentially lines in this range.
}
\label{fig:fig5}
\end{figure}

Several exciting applications of this observational capability are illustrated in Figure \ref{fig:fig5}.  With a wide RF band, one will be able to observe all four key dense gas tracers, their carbon isotopologues, and their deuterated counterparts simultaneously (Figure \ref{fig:fig5}, tuning shown in blue).  The wide RF and IF bands can also probe all four main carbon and oxygen isotopologues of CO in one wideband spectral setup covering 109.5–115.3~GHz (and more), leaving the LSB free to probe important molecular transitions at lower frequencies (e.g. HCO$^+$, N$_2$H$^+$, HCN, HNC, etc.; see \ref{fig:fig5}).
Neither of these tunings would be possible with the narrow RF Band 2 design. To get all the lines within the green tuning in Figure \ref{fig:fig5}, Band 3 would need to be extended to reach IF ranges 4--16~GHz, which may involve a larger IF bandwidth upgrade than planned in the currently ongoing Band 3 upgrade study.

Regarding complex organic molecules (COMs), the wide RF band would allow observations of a majority of COM transitions in 1 or 2 tunings. This is important for species with increasing number of atoms and complexity because to secure their detection, a high number ($>$30) of detected transitions are needed. As an example, the number of low excitation, relatively strong deuterated glycolaldehyde transitions observable in the 67-116 GHz range is $\sim$160, of which only $\sim$70 are observable in the narrow RF Band 2 range (67--95~GHz), and $\sim$130 in the current Band 3 range (84--116~GHz). To cover the maximum number of transitions with the narrow RF Band 2 + Band 3 would require 5 tunings, compared to only 2 with the wide RF Band 2 (see Figure~\ref{fig:tunings}). Even in a single tuning with the wide RF Band 2, $\sim$80 transitions could be observed simultaneously.

A similar example is the search for the strongest transitions of glycine, the simplest amino acid, which are separated by $\sim$7~GHz \citep{Jimenez-Serra2014}.  Similar to the case for hydrogen recombination lines below (Figure \ref{fig:continuum_spectral_setup}), multiple transitions of glycine could be observed simultaneously with the wide RF Band 2.

In general, we conclude this section by noting that for any spectral observations requiring wide frequency coverage to cover multiple lines, the wide RF band can reduce integration times dramatically, even if the  observing time at a single frequency would be 10\% higher.  

\subsection{Continuum Sensitivity}

Of the 4768 observations in Band 3 performed by mid-2017, 3224 (about two thirds) have included at least one wideband ($>$1.8~GHz) spectral window.  This indicates the scientific importance of good continuum measurements in this band, especially in case of very chemically rich sources. 
The upgrade to wideband receivers will offer improve continuum sensitivity in the same integration time, or reduce the integration times by a factor of two if the continuum sensitivity is the primary goal; more generally it will provide much higher flexibility in choosing the line and continuum spectral bands.

\begin{figure}[htb]
\centering
\includegraphics[width=0.8\textwidth]{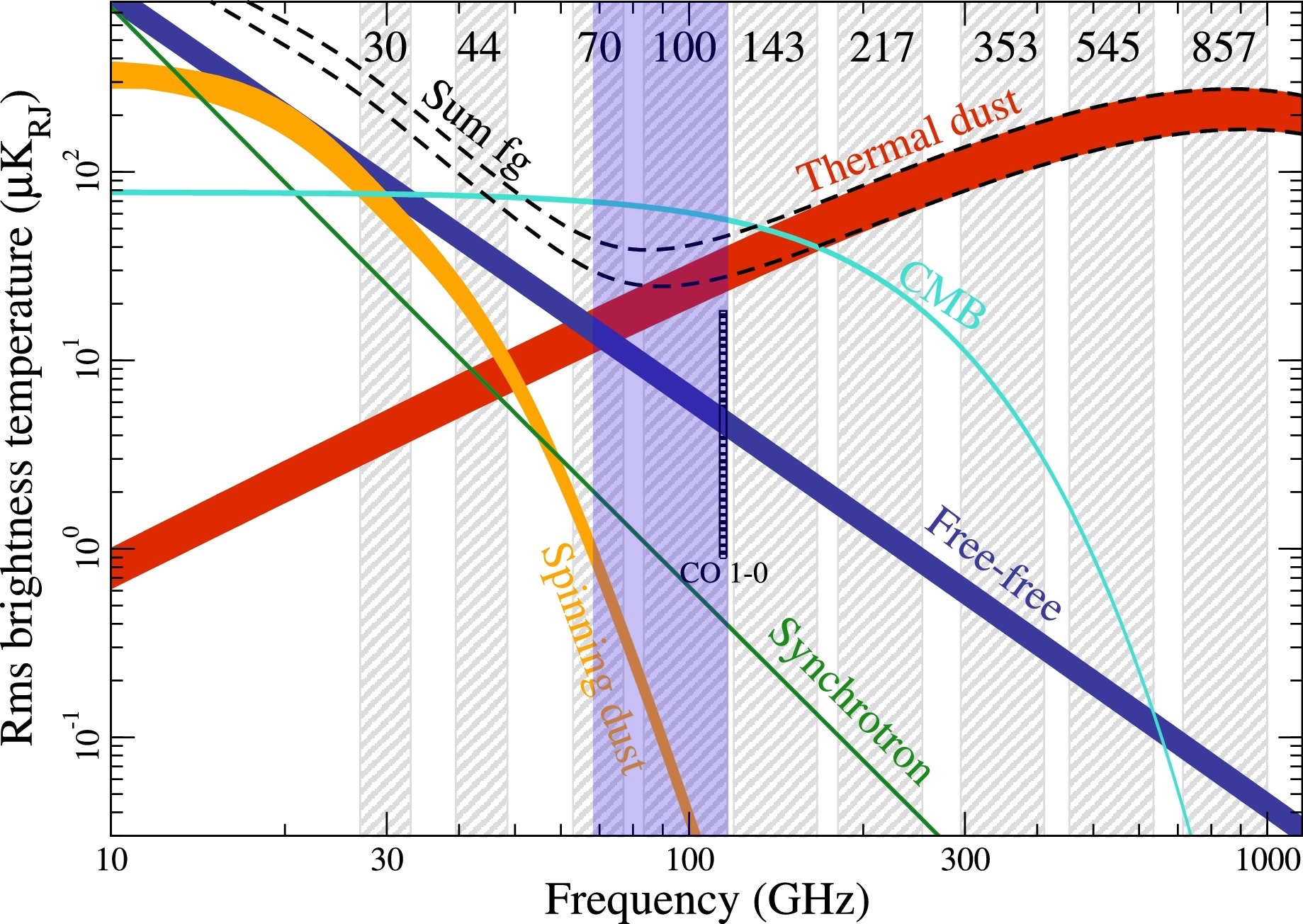}
\caption{\footnotesize 
Combined mm/submm spectrum of foreground dust, synchrotron, anomalous microwave emission, etc, plotted with the band centres of the instruments on the {\it Planck} satellite denoted (grey vertical bands).  
The ALMA wide RF Band 2 (dark blue shaded region) will cover most of the Planck 70 and 100~GHz channels, where the sum of the contributions is lowest. 
Figure is from \citet{Dickinson2018}.
}
\label{fig:ame}
\end{figure}

\begin{figure}[htb]
\centering
\includegraphics[width=0.7\textwidth]{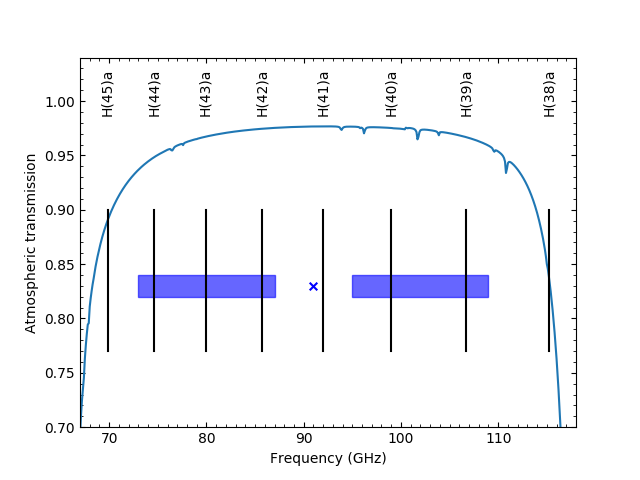}
\caption{\footnotesize Optimal continuum tuning with respect to atmospheric transparency for the wide RF band 2 receiver. The 4--18~GHz IF bandwidth also provides 5 hydrogen recombination lines for free. 
}
\label{fig:continuum_spectral_setup}
\end{figure}

Figure~\ref{fig:continuum_spectral_setup} shows the optimal spectral setup for a pure continuum observation with the wide RF band 2 receiver with 4-18 GHz IF coverage (74-88 GHz + 96-110 GHz). This tuning also contains 5 hydrogen recombination lines, but mostly optimises the atmospheric transparency (see Figure \ref{fig:sky}). Having a wide simultaneous frequency coverage (up to 36 GHz between the tuning edges) provides an improvement of more than thrice what is possible with the current 4-8 GHz IF system, if the digitizers and correlator can handle such bandwidths. Such larger spectral baseline is important for studies of spectral indices, where Bands 2 and 3 are key, as they cover the transition region between synchrotron, free-free and thermal dust dominated continuum processes (see Figure \ref{fig:ame}). 
Note that one only needs an upgrade of the digitisers to get improved spectral indices, this does not necessarily require a new correlator and backends. 

With only 28~GHz of total bandwidth available, and an 8~GHz gap between the lower and upper side band (LSB and USB), the narrow RF Band 2 design has a very limited range of possible broadband continuum setups (spanning only 4~GHz of LO frequencies).  Due to the higher atmospheric noise below 71~GHz, the `optimal' continuum setup for the narrow RF Band 2 would likely favour LSB=71-79~GHz and USB=87-95~GHz (corresponding to an LO frequency of 83~GHz).   

\subsection{Improved {\it uv}-Coverage}

\begin{figure}[htb]
\includegraphics[width=0.329\textwidth]{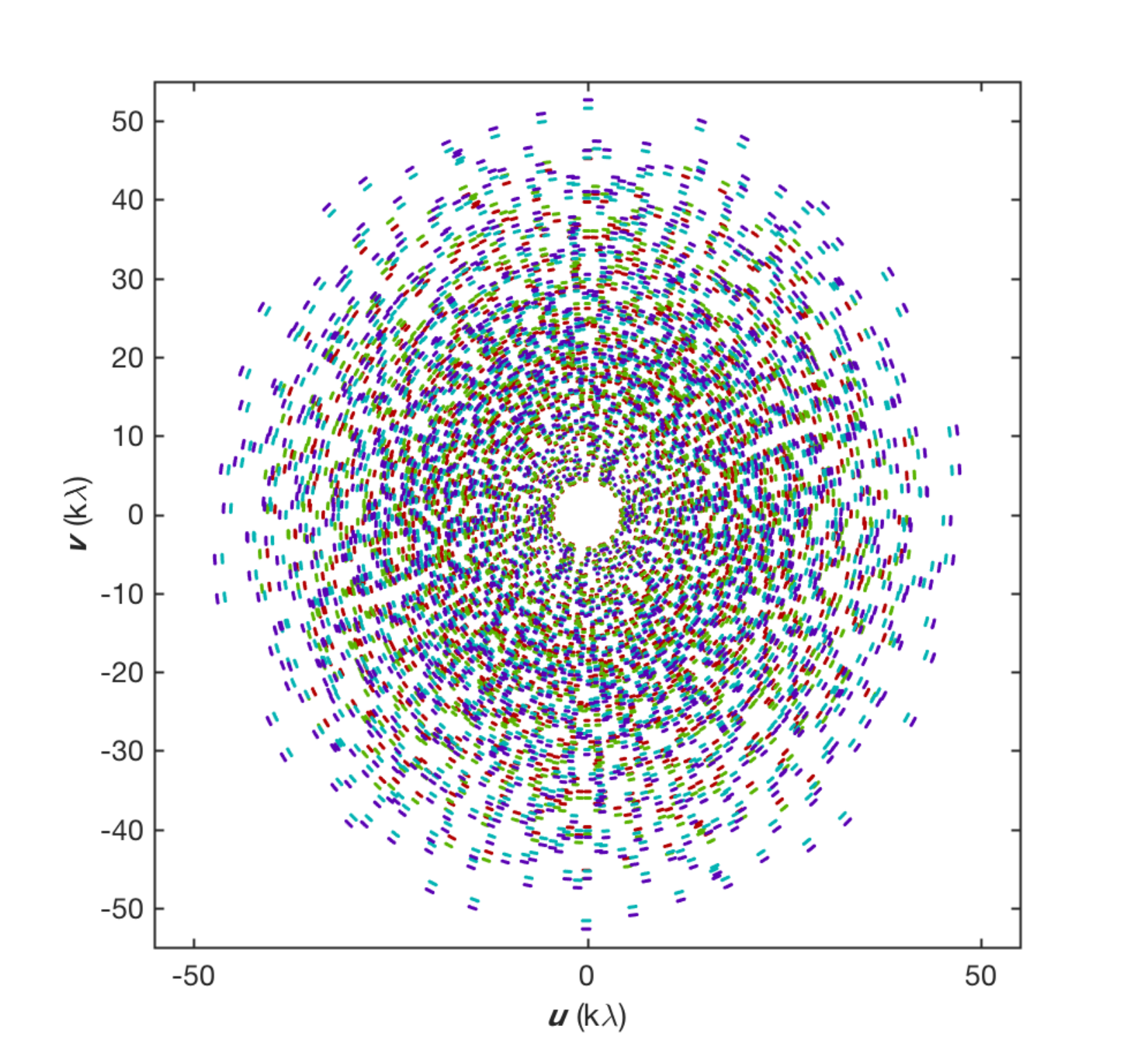}
\includegraphics[width=0.329\textwidth]{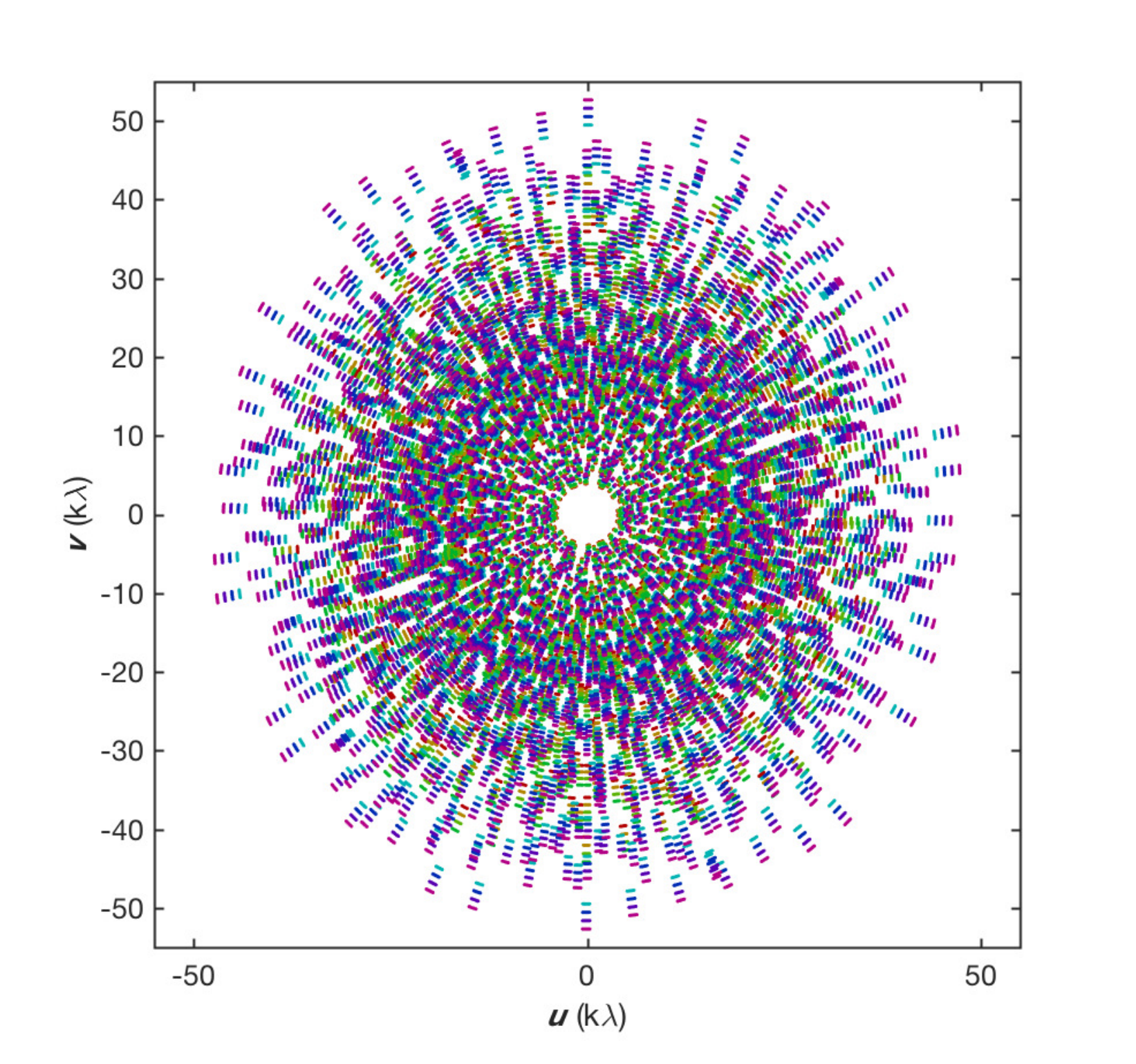}
\includegraphics[width=0.329\textwidth]{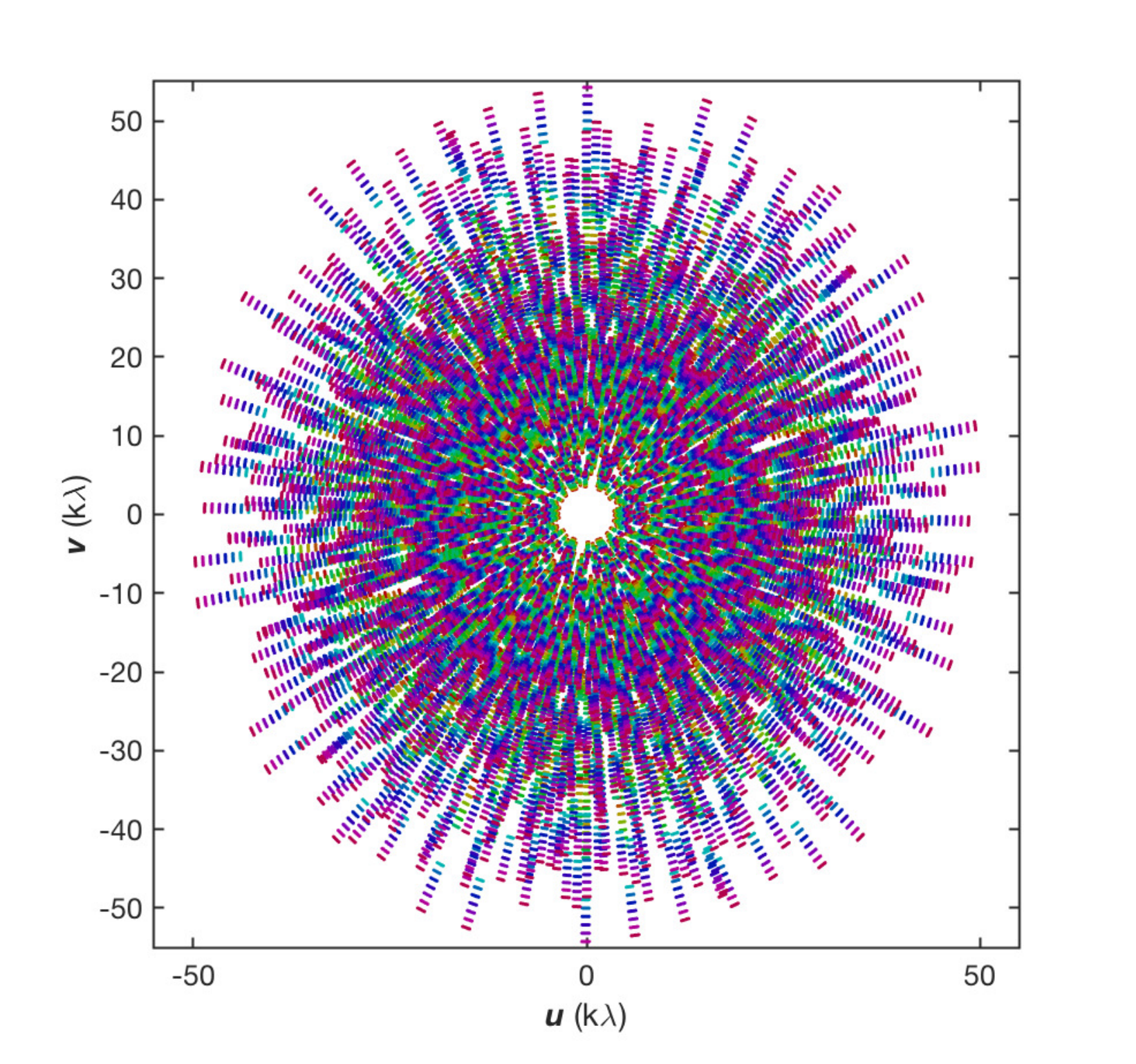}
\caption{\footnotesize The left panel shows the continuum {\it uv}-coverage of a 10-minute track using the original Band 3 with 8~GHz total bandwidth (4~GHz per sideband).  
The middle panel shows the {\it uv}-coverage with narrow RF Band 2, after upgrades to 16~GHz bandwidth (8~GHz per sideband).
The right panel shows the {\it uv}-coverage after upgrades to 28~GHz bandwidth (14~GHz per sideband). Each assumes the most compact configuration of the ALMA 12-meter array.  The wider bandwidth improves the instantaneous {\it uv}-coverage, while the broader frequency coverage simultaneously recovers larger scales and smaller features. Each colour represents a 2~GHz bin in frequency (note: such binning, while chosen for plotting purposes to give the same visual weight to each spectral window, is common when the goal is to maximise continuum imaging sensitivity).
}
\label{fig:uvcoverage}
\end{figure}

The ``best'' continuum range discussed above for wide RF Band 2 will also offer a key advantage in terms of the imaging performance, especially when compared to ALMA's current capabilities. Band 2 will access moderately larger angular scales than Band 3 alone can probe, and with more complete {\it uv}-coverage for multi-frequency continuum imaging of extended sources.  Observations of e.g.\ the Sunyaev-Zeldovich effect from galaxy clusters or extended/complex regions of synchrotron, dust, spinning dust, or free-free emission will benefit from the broad frequency and {\it uv}-coverage simultaneously.  
We note, however, that for sources with complex spectra, the full benefit of the improved {\it uv}-coverage may be difficult to achieve in imaging.
Figure \ref{fig:uvcoverage} shows the {\it uv}-coverage of current 2$\times$4~GHz wide band observations versus those possible with the wide RF \& IF Band 2 coupled to an upgraded 2$\times$14~GHz backend.

\subsection{Transient and time-critical phenomena}

Line ratios in, and instantaneous spectral indices of, transient objects and phenomena, such as objects in our Solar System and gamma-ray bursts, or events such as solar and other stellar flares, will be more accessible with the wide RF band.  In wide band mode, observations will span 36~GHz on sky (with an 8~GHz gap between the USB and LSB).  This will provide a significant lever arm (20-30\% of the central frequency).  The broad instantaneous coverage could also lead to improved imaging and detection of transient events. 

An intriguing possibility for ALMA in the 67-116~GHz range could be fast sweeps or high-cadence spectral observations without switching receivers (i.e.\ avoiding overheads related to receiver biasing, re-positioning the secondary mirror, or pointing corrections).  As new observational modes are continuously being commissioned on ALMA, the wide RF bandwidth may enable relatively fast spectral sweeps over nearly an octave in frequency.  For example, in solar observations, frequency correlates with depth into the solar atmosphere.  As discussed in \S \ref{sec:solar}, many solar phenomena occur on short ($\sim$100 second) time scales, where switching receivers mid-observation would impose prohibitive overheads, potentially ruling out any use of a combination of 2 separate receivers for such observations. Similar considerations apply for comets, where fast rotation periods require efficient observations of molecular lines that may change on timescales $<$1 hour.

\section{Other considerations for Band 2}

\subsection{Partial Redundancy with Band 3}

Both the narrow and wide RF Band 2 version cover part or all of the current Band 3 receivers. Such overlap of receivers may sound like a useless redundancy at first, but it offers several important advantages, both in observing efficiency and technical operations, as also mentioned by \citet{iguchi2018}. To be useful in the current Band 3 range, the Band 2 receiver should have at least similar $T_{\rm rx}$ as the current Band 3 receivers in the spectral range covered by both receivers, see \S~\ref{sec:metric}.

\subsubsection{The 3~mm atmospheric window as a workhorse band for ALMA}

Since the start of ALMA operations, the Band 3 receiver has played a key role in array calibrations, science observations, and is currently the default instrument used in poor weather conditions (PWV$>$4mm), which statistically occur more than half of the available observing time during the summer months at Chajnantor (Otarola et al.\ 2019, PSAP in press).  As of mid January 2019, as many as 7237 of the 31038 observations (23.3\%) in the ALMA science archive were taken in Band 3.\footnote{It is worth noting that while 23.3\% of the observations {\it by number} were in Band 3, $\approx 30\%$ of science integration time was spent in Band 3 in recent cycles.}  
It is difficult to predict how the demand for Band 3 observations will shift once Bands 1 and 2 are both available; however, it is clear that even a reduction in demand by a factor of 2 will still imply that a significant fraction of ALMA’s science will continue to be performed in Band 3.

All but the low frequency edge of Band 2 will be a similar workhorse band for ALMA, in particular during the highest PWV conditions. The wide RF Band 2 receiver has the advantage that this critical frequency range would be covered by two receivers. The frequency overlap would thus reduce the number of band changes and additional calibration overheads this implies, thereby increasing the overall observing efficiency. 
It would also allow the receiver integration engineering team to execute a possible Band 3 upgrade project faster and more efficiently, as the wide RF Band 2 receiver could temporarily take over the observing capabilities in this critical frequency range.

\subsubsection{Very Long Baseline Interferometry}

The 3~mm atmospheric window is popular for mm-wave VLBI.  Many of the systems participating in 3~mm VLBI are being upgraded to record and correlate at least 16~GHz of contiguous bandwidth.  With the wide RF Band 2, ALMA could ensure future compatibility with the worldwide VLBI networks (before an upgraded Band 3 is online, or during its upgrade).

\subsection{Next-Generation ALMA Technologies in the Wide RF Band 2 Receiver}

The wide RF Band 2 system may be the first ALMA receiver to use silicon lens technology, which members of the community studying the Cosmic Microwave Background (CMB) have demonstrated to have low dielectric loss \citep{Parshin1995,Chesmore2018}.  These lenses can be implemented with a high-performance, broadband anti-reflective (AR) surface exhibiting low cross-polarisation leakage \citep{Datta2013}. A prototype Si lens covering the full Band 2 \& 3 frequency range has been integrated into the prototype.  Such lenses may be key in future receiver upgrades where the cryostat opening would otherwise truncate the beam (e.g. at higher frequencies where focal plane arrays are being studied, or for lower frequencies where the telescope optics limit the beam).

\section{Concluding Remarks}

ALMA will soon to be equipped with all of its originally planned receiver bands. All of these meet or exceed the design specifications, contributing to the success of ALMA. However, many other observatories have also benefited from the technological advances driven by ALMA, and have now going beyond this, for example in terms of IF bandwidth. This is recognised by the ALMA 2030 development roadmap, which sets ambitious new science goals driving the next decade of development. Being the last of the originally planned ALMA receiver bands, Band 2 offers unique  opportunities to start implementing these new specifications. In this memo, we have highlighted the scientific advantages of moving towards an IF bandwidth of 4--18~GHz over the full RF range 67--116~GHz. Such an ambitious wideband receiver would be an ideal opportunity to start harvesting the fruits of future ALMA developments, for example in the range of wideband signal processing. We are expecting that this Band 2  receiver could remain state-of-the art well into the 2030's.


\begin{appendix}

\section{Appendix on the Science Case Summary}
\label{sec:appendix}

In this appendix, we highlight the science drivers for the deployment of a wideband receiver system covering the full 67-116 GHz atmospheric window. 
This appendix provides a summary of the most compelling science cases that an ALMA wide RF Band 2 receiver system will enable, discussed in \cite{Beltran2015} and \cite{Fuller2016} in greater detail. It is intended to be part of the documentation package supporting the proposal for the implementation of a wide RF band 2 receiver system.

The wideband system will uniquely enable a broad range of highly compelling science goals, ranging from extragalactic molecular line studies, to detection of complex organic and potentially pre-biotic molecules, characterisation of the fractionation of elements in our own Solar System, evolved stars and young solar analogues, as well as the study of the Solar atmosphere. 

The wide RF band 2 system will enable science not possible with narrow RF Band 2 and Band 3 receivers, as specified in the original project definition from the late 1990, such as where simultaneous measurements are required.  In addition, the wide RF band will reduce the required observing time by a factor of at least two for some science cases and as well as reducing calibration and passband uncertainties which might otherwise compromise science objectives. 


\subsection{Extragalactic spectral surveys}

The first key science driver mentioned in the ALMA Development Roadmap is the origins of galaxies, particularly at $z>2$.  A wide RF Band 2 will be able to probe one or more CO transitions for galaxies at all redshifts $z>2$, as well as opening up the redshift deserts at $0.37< z < 0.72$ and $1.74< z <2$.

ALMA is a powerful telescope to determine spectroscopic redshifts of sources that are obscured in the optical. Since the start of operations, the workhorse band for this work has been Band 3, where at least one CO and/or [CI] line is observable at $z>2$. However, due to the restricted frequency coverage, many unambiguous redshifts remain, which required follow-up in other frequency bands. The contiguous frequency coverage that the wide RF band provides solves this problem. With 28~GHz of IF bandwidth, it will improve the efficiency of this technique by a factor of $>3.5$ in observing time, as compared to the current ALMA system.  At this time, it is not clear what IF bandwidth is being considered for a potential Band 3 upgrade.

\begin{figure}[htb]
\centering
\includegraphics[width=0.8\textwidth]{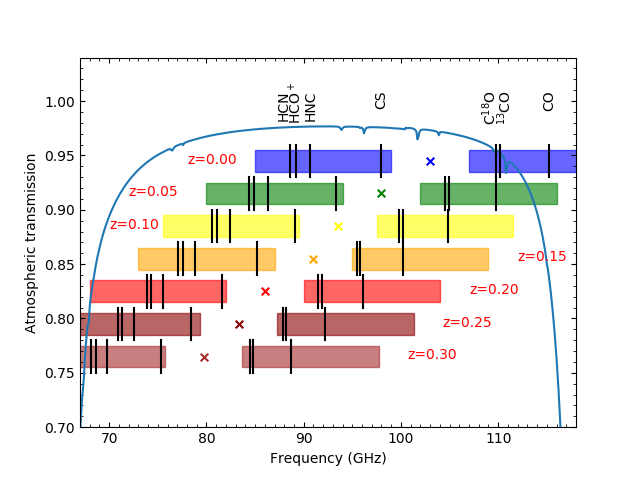}
\caption{\footnotesize 
Configurations which allow the simultaneous observation of a low density tracers ($^{12}$CO(1-0), $^{13}$CO(1-0) and C$^{18}$O(1-0))  and tracers of dense gas (CS(2-1), C$^{34}$S(2-1), $^{13}$CS(2-1), N$^{2}$H+(1-0), HNC(1-0), HCO+(1-0) and HCN(1-0)) for a range of redshifts. For $z>0.05$ at least two dense gas tracers are observable together with CO. At $z>0.15$ all four dense gas tracers are simultaneously observable.   
}
\label{fig:fig6}
\end{figure}

In addition to redshift determinations and spectral line surveys, it will also enable to observe unique combinations of molecular lines in the Local Universe and up to $z\sim0.3$. Simultaneous observations of CO(1-0), $^{13}$CO(1-0) and C$^{18}$O(1-0), CS(2-1), C$^{34}$S(2-1), $^{13}$CS(2-1), and N$^{2}$H+(1-0), and from $z\sim0.2$, in addition HCN(1-0), HNC(1-0) and HCO+(1-0). This will enable simultaneous observations of diffuse and dense gas tracers in galaxies (Figure \ref{fig:fig6}). In the redshift range from $0.1<z<0.25$, such unique setup will only be possible with the wide RF Band 2. 

\subsection{Galactic spectral surveys}

Full spectral line surveys of galactic sources are a key science topic of ALMA. Spectral scans with Band 3 have been approved at high priority for a variety of sources over the past several cycles. ALMA, equipped with a wide RF Band 2 cartridge and with upgraded backends and correlator, will deliver this science for large samples of sources and in a fraction of the observing time currently needed. In addition, below we mention two specific science cases which will benefit from specific lines setup enabled by the wideband systems.

\subsubsection{Complex Organic Molecules in Hot Cores and Hot Corinos}

Hot molecular cores and hot corinos are the most important gas-phase reservoirs of complex organic molecules (COMs), including key species for prebiotic processes, in the Galaxy. COMs in high- and low-mass star-forming regions are regularly observed at (sub)millimeter wavelengths but are severely affected by line blending, especially in turbulent regions where the line widths are broader. The density of rotational transitions of COMs in the submm spectrum decreases with decreasing frequency and their excitation energies are on average lower, especially for heavy molecules. This implies that COMs observed in lower ALMA bands are less blended with other species and their lower energy transitions are more easily excited, which is important when taking into account the low abundance of COMs. Moreover, at high frequency the dust opacity may produce strong absorption of the lines, preventing in some cases the detection of the line emission itself. This effect is much lower in Bands 2 \& 3 than in, e.g., Bands 6 or 7. Therefore, the combination of low dust opacity and blending makes the 3~mm atmospheric  window crucial to observe COMs, and in particular prebiotic ones.  The advantage of a wide RF Band 2 with respect to a narrow RF Band 2 or Band 3 is the number of transitions that can be observed simultaneously or with fewer tunings. This is important for COMs with increasing number of atoms and complexity because to secure their detection, a high number ($>$ 30) of detected transitions are needed. 

\subsubsection{Isotopic ratios and fractionation}

The third key science driver in the 2030 ALMA Development Roadmap is the origins of planets. In addition to improved continuum imaging of the dust in protoplanetary disks, a wide RF Band 2 can help address this through studies of the isotope ratios, CO snowline, and fractionation.

Deuterium fractionation provides an extremely sensitive probe of the evolution of interstellar gas through its coldest, densest phase immediately prior to the formation of a protostar as well as the CO snowline in protoplanetary disks. The fractionation can result in abundance enhancements of multiply deuterated species of a factor of $10^{13}$ above the cosmic deuterium to hydrogen ratio. The wide RF/IF Band 2 will be the most powerful band for deuteration studies: in a single observation, it will be able to observe the four main dense gas species, their deuterium and C/N isotopologues, measuring the deuteration fraction in all four species simultaneously. Similarly, the wide RF Band 2 will simultaneously be able to measure the fractionation of carbon ($^{13}$C), nitrogen ($^{15}$N), oxygen ($^{17}$O and $^{18}$O), and sulphur ($^{33}$S and $^{34}$S), probing of the stellar processing of the interstellar medium and the distribution of the mass of the stars responsible for this processing. 

\begin{figure}[htb]
\centering
\includegraphics[width=0.5\textwidth]{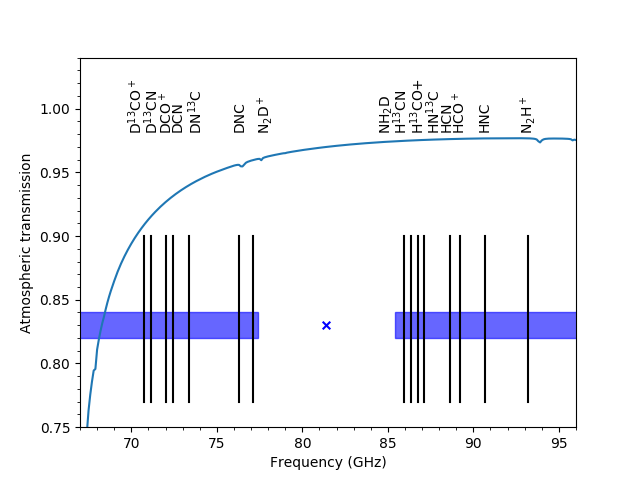}\includegraphics[width=0.5\textwidth]{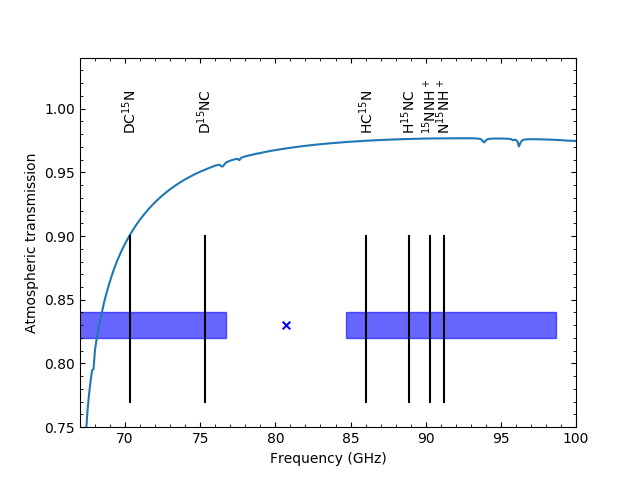}
\caption{\footnotesize 
A single LO setting allows the observation of the important dense gas tracers (HCN, HNC, HCO$^+$ and N$_{2}$H$^+$) and their deuterium isotopologues as well as their $^{13}$C substituted species (Left). This setting will allow the imaging of the deuteration fraction in all these species simultaneously. This same setting will also provide a measurement of the $^{15}$N fractionation in these same species (Right).
Only selected lines are shown here.
}
\label{fig:deut}
\end{figure}

\subsection{Evolved stars}

\begin{figure}[tbh]
\centering
\includegraphics[width=0.9\textwidth]{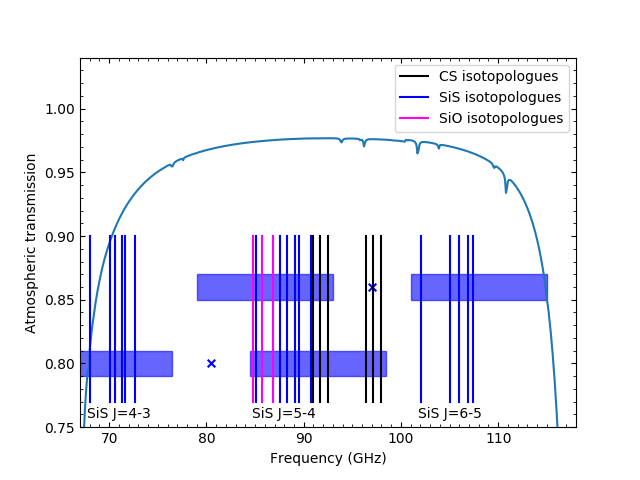}
\caption{
\footnotesize 
Two possible frequency settings optimised for studying the isotopic fractionation in silicon, oxygen, sulphur and carbon as probes of convection and nuclear processing in late stage stellar evolution. The settings cover a range of isotopologues of SiO, SiS and CS with the different settings observing a different pair of SiS transitions.}
\label{fig:fract}
\end{figure}

Isotopic fractionation in evolved stars and the yields to the interstellar medium for the cosmic chemical evolution of galaxies features prominently in ALMA science and the science cases for Band 2. A wide RF Band 2 offers the possibility of effectively carry on a number of fractionation studies with setups that will be particularly useful to constrain the fractionation of O, C, S and Si. One particularly interesting setup, only allowed by the wide RF Band 2 design, could cover the SiO, SiS and CS isotopologues simultaneously (see Figure \ref{fig:fract}).

\subsection{Dust properties and evolution}
\label{sec:dust_cont}

The slope of the emission flux density as a function of wavelength traces the dust grain size distribution, becoming shallower as the grains become larger. By measuring the variation in this spectral slope across the (sub)mm wavelength window --  from the pristine interstellar medium to protoplanetary disks -- it is possible to trace the evolution of the interstellar dust as it grows on the first steps to form the rocky cores of planets. With its wide frequency coverage combined with the higher precision calibration than could be obtained with separate observations, the wide RF Band 2 will be essential in providing high fidelity constraints on grain growth as well as the separation of the emission from small spinning dust particles and the presence of cm-sized pebbles. For this purpose a setup such as that shown in Figure \ref{fig:sky}, located within the best transparency region of the 67-116 GHz window, will be particularly well-suited for continuum observations.

\subsection{High Resolution Studies of the Sunyaev-Zeldovich Effect}
\label{sec:sz}

 The thermal Sunyaev-Zeldovich (SZ) effect probes the line of sight electron pressure in hot gas ($>10^7$ K) such as that in groups and clusters of galaxies \citep{Mroczkowski2019}.  At the level of resolution and sensitivity of ALMA, observations of the SZ effect serve as a powerful complement to X-ray observations of the same gas \citep{DiMascolo2019}, and are expected to be able to probe the SZ effect from individual massive galaxies in the near future.  While the intensity of the SZ decrement peaks in ALMA Band 4, considerations of atmospheric noise contribution position the wide RF Band 2 to be significantly more sensitive.

Band 2 offers two relatively clean windows, from 78-85~GHz and 99-109~GHz, which are excellent for probing the SZ decrement due to the SZ effect.  These windows are largely free of (foreground) local molecular line emission (see Figure \ref{fig:ame}), and sit near the minimum in contamination from dusty thermal and radio sources (see \S \ref{sec:dust_cont}).

\ignore{
\begin{figure}[htb]
\centering
\includegraphics[width=0.7\textwidth]{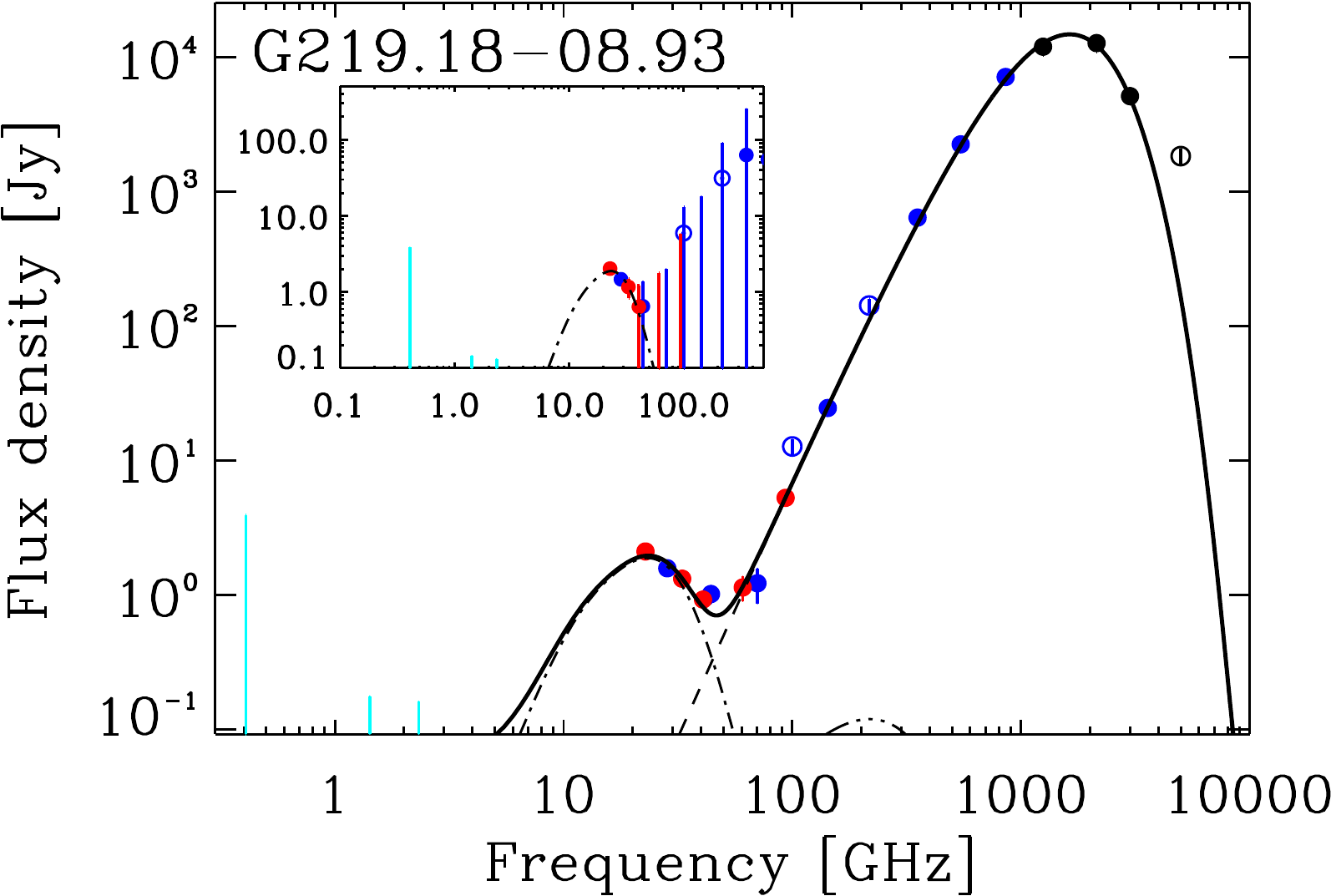}
\caption{\footnotesize 
Bands 2 \& 3 offer a relatively clean window, where dust and synchrotron emission are roughly comparable.  This is important for reducing foreground contamination in studies of, e.g., the SZ effect.}
\label{fig:galdust}
\end{figure}
}

\subsection{Planetary and Solar System}
\label{sec:protoplan}

Comets provide direct samples of the primitive chemical composition of the early solar nebula. The deuteration fraction of cometary material helps tracing the delivery of water to the planets and their satellites, while the fractionation of other species constrains the initial seeding of the proto-solar cloud and hence the environment in which the solar system formed. Since the emission from comets varies on short time scales during their transit though the inner solar system, simultaneous observations of all the lines necessary to measure the fraction are crucial. The wide RF Band 2 is the most efficient band for such observations. 

\subsection{Solar Observations}
\label{sec:solar}

\begin{figure}[htb]
\centering
\includegraphics[width=0.95\textwidth]{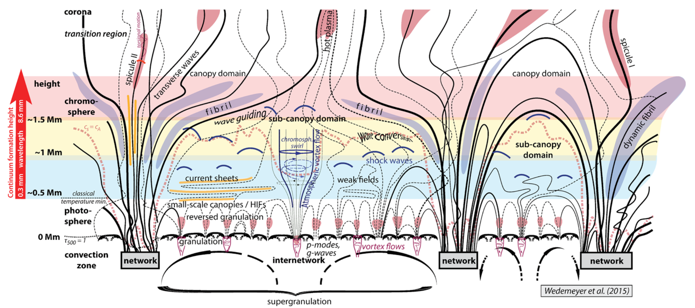}
\caption{\footnotesize 
Regions of the Solar atmosphere probed by millimetre observations. The 67-116~GHz frequency range probes the high chromosphere and, potentially, the transition into the corona (adapted from \citet{Wedemeyer2016}).
}
\label{fig:solar}
\end{figure}

As the millimetre continuum emission height in the Solar atmosphere increases with wavelength, observing at widely spaced frequencies implies sampling different scale heights (see Figure \ref{fig:solar}). The 2.7-4.3~mm wavelength range covered by the wide RF Band 2 design is particularly interesting as it will allow one to probe the high chromosphere and potentially the transition into the solar corona.
Simultaneous sampling of a range of scale heights in the Solar atmosphere thanks to the wide IF bandwidth will have a very large potential for studying the propagation of waves and oscillations, which would allow us to develop helio-seismologic techniques. It would also provide rich data sets for inversion codes which would output 3D time series of the physical properties of the chromospheric plasma. The availability of these powerful and unprecedented observational capabilities would have far-reaching implications for the study of our Sun.

\section{Appendix on LNA performance}
\label{sec:appendixb}

This appendix helps to assess the noise versus bandwidth trade-off in LNA design and implementation. All of the designs presented here are from one group (University of Manchester), using the same modelling tools and fabrication process, but optimised for different applications.  

\begin{figure}[htb]
\begin{center}
\includegraphics[width=0.99\textwidth]{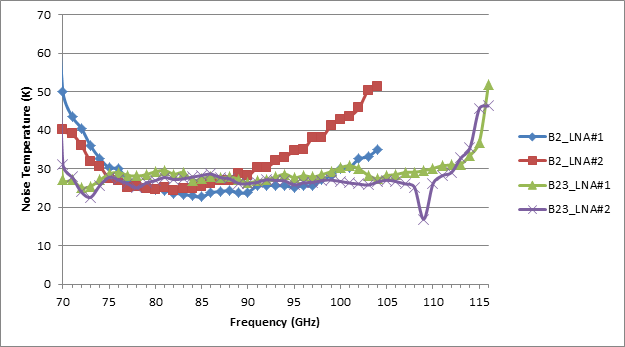}
\end{center}
\caption{\footnotesize Noise temperature performance of two narrowband and two wideband Band 2 LNAs.  The LNAs optimised for performance specifically in the 67-84 GHz range have lower noise temperatures by 2-3~K, demonstrating the metric outlined here (\S \ref{sec:metric}) is attainable now, with state-of-the-art technology.}
\label{fig:lna_perf}
\end{figure}

Figure \ref{fig:lna_perf} shows the noise temperature performance of two narrowband and two wideband Band 2 LNAs.  
The noise temperatures of these LNAs were measured in the cryostat at Caltech at an ambient temperature of 20~K.  The MMICs used in these LNAs are all from the 2015 Northrop Grumman Corporation (NGC) / NASA Jet Propulsion Laboratory (JPL) 35~nm wafer run.  These Band 2 LNAs both use the same MMIC but use slightly different body designs, which results in the variation in frequency performance that can be seen in the figure; B2\_LNA\#1 has better performance higher in the frequency band than B2\_LNA\#2.  Both of the wideband LNAs use the same designs for the MMIC and housing.

Simulations carried out by Manchester also indicate a similar 2-3~Kelvin difference in noise over the 67-84~GHz range for the two designs (with the narrowband-optimised design achieving modestly better performance).

\end{appendix}


\bibliographystyle{aasjournal}
\bibliography{references}

\end{document}